\documentclass[12pt,reqno]{article}
\usepackage{amsmath}

\allowdisplaybreaks

\renewcommand{\i}{\mathrm{i}}
\renewcommand{\k}{\kappa}
\newcommand{\w}{\omega}
\newcommand{\2}{\sqrt{2}}
\newcommand{\al}{\alpha}
\newcommand{\be}{\beta}
\newcommand{\ga}{\gamma}
\newcommand{\de}{\delta}
\newcommand{\la}{\lambda}
\newcommand{\bla}{\lambda^\dag}
\newcommand{\ep}{\varepsilon}
\newcommand{\bpsi}{\psi^\dag}
\newcommand{\bxi}{\xi^\dag}
\newcommand{\si}{\sigma}
\newcommand{\gah}{\hat{\gamma}}
\newcommand{\cA}{\mathcal{A}}

\newcommand{\cC}{\mathcal{C}}
\newcommand{\cF}{\mathcal{F}}
\newcommand{\half}{\tfrac{1}{2}}
\newcommand{\quart}{\tfrac{1}{4}}
\newcommand{\tab}{\quad\,}
\newcommand{\p}{\partial}
\newcommand{\ds}{\partial\!\!\!/}
\newcommand{\Ds}{D\!\!\!\!/}
\newcommand{\ps}{p\mspace{-8mu}/}
\newcommand{\bC}{\bar{C}}
\newcommand{\cD}{\mathcal{D}}
\newcommand{\ol}[1]{\overline{#1}}
\newcommand{\grn}[1]{\langle #1 \rangle}
\newcommand{\e}[2]{e_{#1}{}^{#2}}
\newcommand{\frc}[2]{\frac{\raisebox{-2pt}{$#1$}}{#2}}

\begin{document} 

\begin{flushright} \small
 YITP--SB--01--38 \\ hep-th/0108204
\end{flushright}
\bigskip

\begin{center}
 {\large\bfseries Ward Identities for N=2 Rigid and Local
  Supersymmetry \\[1ex] in Euclidean Space} \\[5mm]
 Ulrich Theis\footnote{Supported by the Deutsche Forschungsgemeinschaft}
 and Peter van~Nieuwenhuizen\footnote{Supported by the National Science
 Foundation under grant no.\ PHY-972 2101} \\[2mm]
 {\small\slshape
 C.N.~Yang Institute for Theoretical Physics \\
 State University of New York at Stony Brook \\
 Stony Brook, NY 11794-3840, USA \\[2pt]
 vannieu,theis@insti.physics.sunysb.edu}
\end{center}
\vspace{5mm}

\hrule\bigskip

\centerline{\bfseries Abstract} \medskip
We construct and check by explicit Feynman diagram calculations the BRST
Ward identities for $N=2$ rigid super Yang-Mills theory and $N=2$
extended supergravity in four-dimensional Euclidean space without
auxiliary fields. We use the Batalin-Vilkovisky formalism. In the
supergravity case we need one new contractible pair of complex spinor
fields to obtain the usual gauge-fixing term and corresponding
Nielsen-Kallosh ghosts.
\bigskip
\hrule
\bigskip

\section{Introduction} 

Many calculations these days of processes involving the AdS/CFT
correspondence and its connection to ten-dimensional supergravity
theories use Euclidean formulations of the field theories instead of
Minkowskian formulations, see for example \cite{AGMOO}. It seems useful
to discuss the properties of these field theories at the quantum level,
in particular to study the Ward identities. The present paper is
intended as a pilot program; we shall consistently make simplifications
when they reduce the amount of algebra and do not change the fundamental
issues we study. For example, we shall evaluate the Ward identities for
extended supergravity in flat space instead of anti-de~Sitter space. Our
final results are therefore easy to understand and check, but the more
realistic cases will be more complicated.

One problem that arises with Euclidean $N=1$ supergravities in $D=10$ is
that there do not exist Majorana-Weyl spinors in $D=10$ Euclidean space,
while the Minkowskian theories use such Majorana-Weyl spinors. One
solution to the problem is to view all fields, bosonic as well as
fermionic, as complex in Euclidean space and to forget about such
reality conditions as Majorana spinors, real self-dual antisymmetric
tensors etc. Since all fields appear ``holomorphically'', i.e., the
(complex) fields but not their complex conjugates appear in the action
and transformation rules, supersymmetry holds equally well in the
complex theory in Euclidean space as in the real theory in Minkowski
space. In particular, the symmetries of fermion bilinears and Fierz
rearrangements of terms quartic in fermion fields needed in the proof
of (rigid or local) supersymmetry only depend on properties of the
charge conjugation matrix, but do not require real spinors \cite{N}.
In this sense Euclidean $N=1$ supersymmetries and supergravities do
exist.

Whether these $N=1$ theories exist at the nonperturbative level, i.e.,
whether the path integral makes sense, is another question which we do
not answer in this paper. For Euclidean path integrals, Gibbons, Hawking
and Perry \cite{GHP} have suggested to make one more Wick-like rotation
on the conformal part of the metric, such that the action in the path
integral becomes bounded from below. It may be that one can also make a
corresponding Wick-like rotation on a conformal part of the gravitino,
in such a way that the new action is still supersymmetric, but we have
not studied this problem. We will study a Ward identity in Euclidean
$N=2$ supergravity, and find that with the standard (unmodified) Feynman
rules this Ward identity in Euclidean space is satisfied.

In this article we consider $N=2$ supersymmetric theories. For them,
one can combine a pair of Majorana spinors into a Dirac spinor, and a
complex Dirac spinor in Euclidean space can also be written as a pair
of Majorana spinors, but with a modified (``symplectic'') Majorana
condition. The real bosonic fields in Minkowski space are again real
in Euclidean space. Hence, for $N=2$ theories complexification is
not needed.

However, there is a second problem with Euclidean theories, namely
certain spinless fields acquire an extra minus sign in front of their
kinetic action, and extra factors of $\i$ in front of their Yukawa
couplings \cite{Z}. The origin of these minus signs and factors of
$\i$ was pointed out in \cite{vNW}, where it was noted that a
pseudoscalar in four dimensions can be represented in terms of four
scalar fields $\phi_i$ as $\ep^{\mu\nu\rho\si}\p_\mu\phi_1\p_\nu\phi_2
\p_\rho\phi_3\p_\si\phi_4$. Precisely one derivative is timelike and
hence the Wick rotation produces a factor $\i$. This mechanism explains
why the kinetic action of the axion in $D=10$ changes sign in Euclidean
space.

In \cite{vNW} the Wick rotation was extended to fermions. The
basic idea was that the Wick rotation for a vector $(A_0,\vec{A}\,)
\rightarrow(\i A_4^E,\vec{A}^E)$ is really an induced representation
$A_\mu(t,\vec{x})\rightarrow O_\mu{}^\nu A_\nu^E(\tau,\vec{x})$ with
$O_\mu{}^\nu=\mathrm{diag}(\i,1,1,1)$ (we denote Minkowski time by
$t$ and Euclidean time by $\tau$). It was then natural to assume
that for fermions $\psi\rightarrow O\psi_E$, where $O$ is again a
$4\times 4$-matrix with spinor indices, but now the matrix will not
be diagonal. Consistency arguments then show that $O$ is a complex
Lorentz rotation in the $t$-$\tau$ plane, which is the spinor
representation of the Wick rotation. With these matrices $O$ for (real
or complex) bosons and complex fermions one can construct from every
field theory in Minkowski spacetime a corresponding theory in Euclidean
space. One finds then that certain internal compact rigid symmetries
become noncompact after the Wick rotation, just like the noncompact
spacetime symmetry SO(3,1) becomes the compact SO(4). This exchange of
compactness and noncompactness between spacetime and internal symmetries
is not surprising if we take the spacetime origin of internal symmetries
provided by the Kaluza-Klein program into account. For example, the
SO(9,1) symmetry of $N=1$, $D=(9,1)$ super Yang-Mills theory becomes an
SO(3,1) $\times$ SO(6) symmetry if one compactifies on a 6-torus, but it
becomes SO(4) $\times$ SO(5,1) with a noncompact internal SO(5,1) if one
compactifies on a torus T(5,1) where the time axis has been compactified
\cite{BT}. The same features arise in the compactification of $N=1$,
$D=(9,1)$ supergravity on $\mathrm{S}^3\times\mathrm{AdS}_3$ leading to
$N=4$, SU(2)$\times$SU(1,1) supergravity in four-dimensional
Euclidean space \cite{V}.

Although some of the properties of Euclidean supersymmetric field
theories have been explained in this way, the next question, dynamics
and Ward identities in Euclidean space, has been less studied. One
reason for this absence of such studies is that for super Yang-Mills
theories in $x$-space the gauge fixing term and the ghost action break
the rigid susy. One must then derive Ward identities for explicit
symmetry breaking \cite{FJvN}. For $3+1$ dimensional susy gauge theories
one can (at least for $N=1$ models) go to superspace with (anti-) ghost
superfields and preserve rigid susy. However, for $D>4$ and $N>1$ no
full-fledged off-shell superspace formalism exists, whereas for $D=4$,
$N=1$ one runs into the problems with Majorana spinors mentioned above.

Over a decade ago, it was noticed \cite{BC} that one can treat rigid
symmetries of Minkowskian field theories on the same footing as local
gauge symmetries, and use a BRST formalism \cite{BRST} (more precisely
a field-antifield formalism, also called a BV-formalism \cite{ZJ,BV})
for all (rigid and local) symmetries simultaneously. This was applied
to rigid supersymmetry in \cite{BBOW,M,MPR}. One of the purposes of this
paper is to apply this formalism to supersymmetric field theories in
Euclidean space and to derive Ward identities. In particular, we shall
analyze in explicit examples whether the minus signs and factors of
$\i$ due to the Wick rotation of pseudoscalars are compatible with
the BRST Ward identities in Euclidean space.

Of course, the BV formalism allows one to deal with locally or globally
supersymmetric theories without auxiliary fields. This is in our case an
enormous advantage, because the auxiliary fields for $N=2$ supergravity
are quite complicated \cite{FV}. For $N=2$ super Yang-Mills theory
\cite{F} they are not complicated and we could have incorporated them.

We shall study $N=2$, $D=4$ Euclidean super Yang-Mills theory. In this
model the two Majorana spinors are written, in Minkowski spacetime and
in Euclidean space, as one complex spinor. This model has one real
scalar and one real pseudoscalar, so it is an ideal testing ground for
our purposes. Applying the Wick rotation rules, one obtains a hermitean
action in Euclidean space, which has again $N=2$ supersymmetry, but with
a kinetic action for the pseudoscalar with the wrong sign. We derive
some Ward identities and explicitly verify them at the one-loop level.
Our conclusion is that the Euclidean Ward identities are satisfied if
one uses the minus sign and factors of $\i$ for the pseudoscalars, but
one can equally well redefine $\varphi\rightarrow\i\varphi$ both in the
action and in the Ward identities, thus obtaining the standard
positive-definite kinetic action, at the expense of hermiticity of the
action. Since unitarity is a concept in Minkowski space only, violation
of hermiticity of a Euclidean action poses no problem.

One is of course also interested in local supersymmetry, i.e.\
supergravity. We consider $N=2$, $D=(3,1)$ supergravity which was the
first extended supergravity, constructed in 1976 \cite{FvN}. It unifies
electromagnetism and gravity, and does so by introducing two gravitinos.
In this model, the two Majorana gravitinos combine in Minkowski
spacetime into one complex gravitino $\psi_\mu^\al$. The Wick rotation
now involves both a matrix $O_\mu{}^\nu$ for the vector index and a
matrix $O^\al{}_\be$ for the spinor index. The other fields, the real
vielbein $\e{\mu}{a}$ and the real vector field $A_\mu$, are not
complexified and are again real in Euclidean space. We construct $N=2$,
$D=4$ Euclidean supergravity including the supersymmetric cosmological
constant by using the Wick rotation of \cite{vNW} with the $O$-matrices.
For example, the Minkowski transformation rule for the gravitino
$\psi_\mu$ with $\mu=0$ reads
 \begin{equation*}
  \de_\epsilon \psi_0 = \ldots + \tfrac{1}{\2} g \k^{-2} \ga_0
  \epsilon\ ,\quad \ga_0 = \e{0}{a} \ga_a\ ,
 \end{equation*}
in anti-de~Sitter supergravity \cite{DF}. It becomes in Euclidean
space\footnote{Use that if $t\rightarrow -\i\tau$, the covariant index
$\mu=0$ of $\e{\mu}{a}$ acquires a factor $+\i$, but the contravariant
index $a=0$ a factor $-\i$. Then use $-\i\ga_0=\ga_4$. Note that in flat
space $\e{\mu}{a}=\de_\mu^a$ both in Minkowski and in Euclidean space.}
 \begin{equation*}
  \i O^\al{}_\be\, \de_\epsilon \psi^{\be\,E}_4 = \ldots + \tfrac{1}{
  \2} g \k^{-2}\, (\i \ga_4 \epsilon)^\al\ ,
 \end{equation*}
and defining $\epsilon\equiv O\epsilon_E$, one obtains
 \begin{equation*}
  \de_\epsilon \psi^E_4 = \ldots + \tfrac{1}{\2} g \k^{-2} (O^{-1}
  \ga_4 O) \epsilon_E\ .
 \end{equation*}
One identifies then $O^{-1}\ga^4 O$ as the Euclidean Dirac matrix
$\ga^4_E$, which, in fact, is equal to the usual (Minkowski) matrix
$\ga^5$ \cite{vNW}. Similarly, in Dirac actions we encounter the
structure $\mathcal{L}=\bpsi\ga_E^5\ga^\mu_E\p_\mu\psi$, where $\mu$
runs from 1 to 4, because $\ga^4$ in $\ol{\psi}=\bpsi\ga^4$ becomes
$O\ga^4 O=\ga^4=-\ga^5_E$. This action clearly preserves SO(4) symmetry
with generators $\quart[\ga_E^\mu,\ga_E^\nu]$.

In the quantization of gravity theories with spinors in Minkowski space,
there are not only Einstein ghosts $C^\mu$ and antighosts $\bC_\mu$, but
also Lorentz ghosts $C^{ab}$ and antighosts $\bC_{ab}$. We can also
first make a Wick rotation on the classical theory, and then quantize.
One obtains then SO(4) (anti-) ghosts. The Lorentz ghosts are, as
expected, related to the SO(4) ghosts by a Wick rotation. We can fix the
Lorentz symmetry at the classical level by requiring the vielbein to be
symmetric, but it is better to avoid such constraints on the vielbein at
the quantum level and to accept the presence of the Lorentz (anti-)
ghosts. If desired, one can eliminate them from the quantum theory by
their algebraic field equations \cite{vN}. (We recall that this does not
mean that we put the theory on-shell, but, rather, that one performs a
simple Gaussian integration in the path integral.) There is no problem
with the negative signs due to the Lorentz metric for Berezin
integration. We keep the Lorentz or SO(4) (anti-) ghosts at all stages.

Since the standard BV formalism deals with gauge theories and the
presence of rigid symmetries requires some extra discussion, and also
since for applications the supergravity part is more interesting than
the super Yang-Mills part, we begin with the former.

\section{Euclidean N=2 Supergravity} 

Pure $N=2$ extended supergravity in four-dimensional Minkowski
spacetime was formulated in \cite{FvN}. The field content consists of
a vielbein $\e{\mu}{a}$, a vector $A_\mu$ (the graviphoton) and two
Majorana spinors $\chi_\mu^1$, $\chi_\mu^2$ (the gravitinos). In
\cite{DF} the model was generalized to include a minimal coupling of
the graviphoton to the gravitinos, which requires the presence of a
non-vanishing cosmological constant. For the continuation to
Euclidean space it is convenient to combine the gravitinos into a
complex Dirac spinor $\psi_\mu=(\chi_\mu^1+\i\chi_\mu^2)/\2$. The
action then reads for the Minkowski theory\footnote{We use the
following conventions in Minkowski spacetime: $\{\ga^a\, ,\ga^b\}=2
\eta^{ab}$, $\eta_{00}=-1$, $\ga^{ab}=\ga^{[a}\ga^{b]}=\half(\ga^a\ga^b
-\ga^b\ga^a)$, $\ol{\psi}=\bpsi\ga^4$, $\ga^4=\i\ga^0$, $\ga_5=\ga^1
\ga^2\ga^3\ga^4$, $\ga^\mu=\ga^a\e{a}{\mu}$, $\ep^{0123}=1$, $e=\det
\e{\mu}{a}$.}
 \begin{align}
  S_0 = \int\! d^4x\ \Big[ & \frc{e}{2\k^2}\, R(e,\w) - \i\,
	\ep^{\mu\nu\rho\si} \ol{\psi}_\mu \ga_5 \ga_\nu \cD_\rho
	\psi_\si - \frc{1}{4}\, e\, F^{\mu\nu} F_{\mu\nu} + \frc{6g^2}{
	\k^4}\, e \notag \\*
  & - \frc{\i\k}{2\2}\, \ol{\psi}_\mu \big[ e\, (F^{\mu\nu} + \cF^{\mu
	\nu}) + \i \ga_5 (\tilde{F}^{\mu\nu} + \tilde{\cF}^{\mu\nu}) +
	\frc{4\i g}{\k^2}\, e\, \ga^{\mu\nu} \big] \psi_\nu \Big]\ ,
 \end{align}
where a long bar denotes Dirac conjugation. (We shall use short bars
as in \eqref{s-antigh} to denote antighosts.) The curvature scalar is
given in terms of the inverse vielbein and the spin connection by
 \begin{equation}
  R = \e{a}{\mu} \e{b}{\nu} R_{\mu\nu}{}^{ab}\ ,\quad R_{\mu\nu}{}^{ab}
  = \p_\mu \w_\nu{}^{ab} - \p_\nu \w_\mu{}^{ab} + \w_\mu{}^{ac}
  \w_{\nu c}{}^b - \w_\nu{}^{ac} \w_{\mu c}{}^b\ ,
 \end{equation}
while $\cF_{\mu\nu}$ and $\tilde{\cF}^{\mu\nu}$ denote the
supercovariant field strength and its Hodge-dual respectively,
 \begin{equation} \label{cF_M}
  \cF_{\mu\nu} = F_{\mu\nu} + \frc{\i\k}{\2}\, (\ol{\psi}_\mu
  \psi_\nu - \ol{\psi}_\nu \psi_\mu)\ ,\quad \tilde{\cF}^{\mu\nu}
  = \half \ep^{\mu\nu\rho\si} \cF_{\rho\si}\ .
 \end{equation}
The operator $\cD_\mu$ entering the kinetic term of the gravitinos is
the Lorentz- and gauge-covariant derivative,
 \begin{gather}
  \cD_\mu \psi_\nu = (D_\mu - \i g A_\mu) \psi_\nu \notag \\[2pt]
  D_\mu \psi_\nu = \p_\mu \psi_\nu + \quart \w_\mu{}^{ab} \ga_{ab}
	\psi_\nu\ ,\quad D_\mu \e{\nu}{a} = \p_\mu \e{\nu}{a} -
	\w_{\mu b} {}^a \e{\nu}{b}\ .
 \end{gather}
The spin connection $\w_\mu{}^{ab}$ is the function of the vielbein and
gravitinos that satisfies its equation of motion $\de S_0/\de\w_\mu
{}^{ab}=0$ (1.5 order formalism \cite{vN2}),
 \begin{equation} \label{w}
  \w_{\mu ab} = \w_{\mu ab}(e) + \frc{\k^2}{4}\, (\ol{\psi}_a \ga_\mu
  \psi_b + \ol{\psi}_a \ga_b \psi_\mu - \ol{\psi}_\mu \ga_b \psi_a -
  a \leftrightarrow b)\ .
 \end{equation}

The supersymmetry transformations with Dirac (i.e., complex) spinor
parameter $\epsilon(x)$ read
 \begin{align}
  \de_\epsilon \e{\mu}{a} & = \half \k (\ol{\epsilon} \ga^a \psi_\mu
	- \ol{\psi}_\mu \ga^a \epsilon) \notag \\[2pt]
  \de_\epsilon A_\mu & = \tfrac{\i}{\2}\, (\ol{\psi}_\mu \epsilon -
	\ol{\epsilon} \psi_\mu) \notag \\[2pt]
  \de_\epsilon \psi_\mu & = \k^{-1} \cD_\mu \epsilon - \tfrac{\i}{2
	\2}\, \ga^\nu (\cF_{\mu\nu} + \i e\, \ga_5 \tilde{\cF}_{\mu
	\nu})\, \epsilon + \tfrac{1}{\2}\, g \k^{-2} \ga_\mu
	\epsilon\ .
 \end{align}
Using these rules one easily verifies that $\cF_{\mu\nu}$ in
\eqref{cF_M} and $\w_\mu{}^{ab}$ in \eqref{w} are supercovariant (in the
variations the $\p_\mu\epsilon$ terms cancel).

We now employ the Wick rotation prescription for Dirac spinors
presented in \cite{vNW},
 \begin{equation} \label{O}
  \psi\, \rightarrow\, O \psi_E\ ,\quad \bpsi\, \rightarrow\, \bpsi_E\,
  O\ ,\quad O = \mathrm{e}^{\ga^4\ga^5\pi/4}\ ,
 \end{equation}
along with the usual rules $t\rightarrow-\i\tau$ and $A_\mu=(A_0,
\vec{A}\,)_\mu\rightarrow(\i A_4^E,\vec{A}^E)_\mu$. The matrix $O$ is
unitary and satisfies $O\ga^4=\ga^4 O^{-1}$. The Euclidean
$\ga$-matrices are related to those in Minkowski spacetime by
 \begin{equation}
  \ga^a_E = O^{-1} \ga^a O \quad\Rightarrow\quad \ga^i_E = \ga^i\
  ,\quad \ga^4_E = \ga^5\ ,\quad \ga^5_E = \ga^1_E \ga^2_E \ga^3_E
  \ga^4_E= - \ga^4\ ,
 \end{equation}
where $a$ now runs from 1 to 4.
In particular, one has $\ol{\psi}\rightarrow -\bpsi_E\ga^5_E O^{-1}$.
These rules yield the following Euclidean action, where we have
included a factor $(-1)$ so that it enters the path integral via
$\exp(-S_0)$ (from now on we drop all subscripts $E$ that denote
Euclidean quantities)
 \begin{align}
  S_0 = \int\! d^4x\ \Big[ & - \frc{e}{2\k^2}\, R(e,\w) - \ep^{\mu
	\nu\rho\si} \bpsi_\mu \ga_\nu \cD_\rho \psi_\si + \frc{1}{4}\,
	e\, F^{\mu\nu} F_{\mu\nu} - \frc{6g^2}{\k^4}\, e \notag \\*
  & - \frc{\i\k}{2\2}\, \bpsi_\mu \ga_5 \big[ e\, (F^{\mu\nu} +
	\cF^{\mu\nu}) + \ga_5 (\tilde{F}^{\mu\nu} + \tilde{\cF}^{\mu
	\nu}) + \frc{4\i g}{\k^2}\, e\, \ga^{\mu\nu} \big] \psi_\nu
	\Big]\ . \label{L_0}
 \end{align}
Note that $S_0$ is hermitean.

The supercovariant field strength and spin connection in Euclidean
space read
 \begin{align}
  \cF_{\mu\nu} & = F_{\mu\nu} - \frc{\i\k}{\2}\, (\bpsi_\mu \ga_5
	\psi_\nu - \bpsi_\nu \ga_5 \psi_\mu) \\[2pt]
  \w_{\mu ab} & = \w_{\mu ab}(e) - \frc{\k^2}{4}\, (\bpsi_a \ga_5
	\ga_\mu \psi_b + \bpsi_a \ga_5 \ga_b \psi_\mu - \bpsi_\mu
	\ga_5 \ga_b \psi_a - a \leftrightarrow b)\ .
 \end{align}
The former occurs in the action and the transformations only in the
combination
 \begin{equation}
  G_{\mu\nu} = \cF_{\mu\nu} + e\, \ga_5 \tilde{\cF}_{\mu\nu}\ .
 \end{equation}
The Wick rotation produces the following local supersymmetry
transformation rules in Euclidean space
 \begin{align}
  \de_\epsilon \e{\mu}{a} & = \half \k (\bpsi_\mu \ga_5 \ga^a \epsilon
	- \epsilon^\dag \ga_5 \ga^a \psi_\mu) \notag \\[2pt]
  \de_\epsilon A_\mu & = \tfrac{\i}{\2}\, (\epsilon^\dag \ga_5 \psi_\mu
	- \bpsi_\mu \ga_5 \epsilon) \notag \\[2pt]
  \de_\epsilon \psi_\mu & = \k^{-1} \cD_\mu \epsilon - \tfrac{\i}{2
	\2}\, \ga^\nu G_{\mu\nu}\, \epsilon + \tfrac{1}{\2}\, g \k^{-2}
	\ga_\mu \epsilon\ .
 \end{align}
Using these rules one can again verify that $\cF_{\mu\nu}$ and $\w_\mu
{}^{ab}$ are supercovariant, and that the action \eqref{L_0} is
invariant.

$S_0$ is also invariant under the following BRST transformations of
the supergravity multiplet\footnote{Replace $\epsilon$ by $-\Lambda\xi$
with imaginary $\Lambda$ and remove $\Lambda$ from the left.}
 \begin{align}
  s_0 \e{\mu}{a} & = C^\nu \p_\nu \e{\mu}{a} + \p_\mu C^\nu \e{\nu}{a}
	- C_b{}^a \e{\mu}{b} + \half \k\, (\bpsi_\mu \ga_5 \ga^a \xi
	- \bxi \ga_5 \ga^a \psi_\mu) \notag \\[2pt]
  s_0 A_\mu & = C^\nu \p_\nu A_\mu + \p_\mu C^\nu A_\nu + \p_\mu C +
	\tfrac{\i}{\2}\, (\bxi \ga_5 \psi_\mu - \bpsi_\mu \ga_5 \xi)
	\notag \\[2pt]
  s_0 \psi_\mu & = C^\nu \p_\nu \psi_\mu + \p_\mu C^\nu \psi_\nu +
	\quart C^{ab} \ga_{ab} \psi_\mu - \k^{-1} \cD_\mu \xi +
	\tfrac{\i}{2\2}\, \ga^\nu G_{\mu\nu} \xi \notag \\*
  & \tab - \tfrac{1}{\2}\, g \k^{-2} \ga_\mu \xi + \i g C\, \psi_\mu
	\notag \\[2pt]
  s_0 \bpsi_\mu & = C^\nu \p_\nu \bpsi_\mu + \p_\mu C^\nu \bpsi_\nu -
	\quart C^{ab} \bpsi_\mu \ga_{ab} + \k^{-1} \cD_\mu \bxi +
	\tfrac{\i}{2\2}\, \bxi G_{\mu\nu} \ga^\nu \notag \\*
  & \tab + \tfrac{1}{\2}\, g \k^{-2} \bxi \ga_\mu - \i g C\,
	\bpsi_\mu\ , \label{s_0-sugra}
 \end{align}
which include general coordinate transformations, local SO(4)
rotations, U(1) gauge transformations and supersymmetry
transformations with corresponding ghosts $C^\mu$, $C^{ab}$, $C$
and $\xi$, $\bxi$ respectively. The latter are commuting Dirac
spinors. The BRST transformations of the ghosts follow from the
commutator algebra of the local symmetries; alternatively, they can
be derived from imposing (on-shell) nilpotency of $s_0$ on the above
fields. We obtain in either case
 \begin{align}
  s_0 C & = C^\mu \p_\mu C + \tfrac{\i}{\2}\, \k^{-1} \bxi \ga_5 \xi
	+ \half A_\mu\, \bxi \ga_5 \ga^\mu \xi \notag \\[2pt]
  s_0 C^\mu & = C^\nu \p_\nu C^\mu - \half\, \bxi \ga_5 \ga^\mu \xi
	\notag \\[2pt]
  s_0 C^{ab} & = C^\mu \p_\mu C^{ab} + C^{ac} C_c{}^b - \half\, \bxi
	\ga_5 ( \ga^\mu \w_\mu{}^{ab} + \tfrac{\i}{\2} \k\, G^{ab}
	+ \2\, g \k^{-1} \ga^{ab}) \xi \notag \\[2pt]
  s_0 \xi & = C^\mu \p_\mu \xi + \quart C^{ab} \ga_{ab} \xi - \half
	\k\, \psi_\mu\, (\bxi \ga_5 \ga^\mu \xi) + \i g C\, \xi \notag
	\\[2pt]
  s_0 \bxi & = C^\mu \p_\mu \bxi - \quart C^{ab} \bxi \ga_{ab} + \half
	\k\, (\bxi \ga_5 \ga^\mu \xi) \bpsi_\mu - \i g C\, \bxi\ .
	\label{s_0-ghosts}
 \end{align}
The ghosts behave like tensors under coordinate transformations, except
for the coordinate ghosts themselves, for which $dC^\mu$ transform as
coordinate vectors \cite{TvN}.

Since we did not include auxiliary fields in the supergravity multiplet,
the commutator algebra closes only on-shell on the gravitinos, which
implies that on these $s_0$ is not nilpotent but squares into a trivial
gauge transformation (namely a field equation). Also one finds that
nilpotency holds on the SO(4) rotation ghosts only modulo the field
equations of the gravitinos,
 \begin{gather}
  s_0^2\, \psi_\mu = M_{\mu\nu}\, \frc{\de S_0}{\de\bpsi_\nu} +
	\frc{\de S_0}{\de\psi_\nu}\, N_{\nu\mu}\ ,\quad s_0^2\, C^{ab}
	= \Sigma^{ab}_\mu{}^\dag\, \frc{\de S_0}{\de\bpsi_\mu} +
	\frc{\de S_0}{\de\psi_\mu}\, \Sigma^{ab}_\mu \notag \\
  s_0^2\, \text{(other fields)} = 0\ ,
 \end{gather}
where
 \begin{align}
  M_{\mu\nu}{}^\al{}_\be & = - \frc{1}{8e}\, \big[ (\ga_5 \ga^\rho \xi
	)^\al (\bxi \ga_5 \ga_\nu \ga_{\mu\rho})_\be - (\ga_5 \xi)^\al
	(\bxi \ga_5 \ga_\nu \ga_\mu)_\be - (\ga_5 \ga_{\mu\nu} \xi)^\al
	(\bxi \ga_5)_\be \notag \\
  & \mspace{65mu} + (\ga_5 \ga^\rho \xi)^\al (\bxi \ga_5 \ga_\rho
	\ga_\nu \ga_\mu)_\be + 2 \bxi \xi\, (\ga_\nu \ga_\mu)^\al{}_\be
	\big] \notag \\[2pt]
  N_{\nu\mu}{}^{\be\al} & = - \frc{1}{8e}\, \big[ (\ga_5 \ga_{\mu\rho}
	\ga_\nu \xi)^\be (\ga_5 \ga^\rho \xi)^\al + (\ga_5 \xi)^\be
	(\ga_5 \ga_{\mu\nu} \xi)^\al + (\ga_\mu \ga_\nu \xi)^\be
	\xi^\al \big] \notag \\[2pt]
  \Sigma_{\mu ab}^\al & = \frc{\k}{8e}\, \e{a}{\rho} \e{b}{\si}\, \big[
	e\, \ep_{\mu\nu\rho\si} (\ga_5 \xi)^\al\, \bxi \ga_5 \ga^\nu \xi
	+ (\ga_{\rho\si} \ga_\mu \ga_5 \xi)^\al\, \bxi \xi \big]\ .
 \end{align}

In order to obtain a nilpotent BRST operator and to quantize the model,
we employ the Batalin-Vilkovisky approach \cite{BV,ZJ}. There exist
off-shell formulations for $N=2$ extended supergravity \cite{FV}, but
these are quite complicated, and working with the BV formalism without
auxiliary fields is much simpler for our aim of checking Ward
identities. With auxiliary fields the local gauge algebra of general
coordinate transformations, local Lorentz (SO(4) in the Euclidean case)
transformations, local supersymmetry transformations and U(1) gauge
transformations closes off-shell. In particular the commutator of two
local susy transformations produces a local Lorentz transformation with
a parameter that depends on the auxiliary fields. This explains why
$s_0^2C^{ab}$ is nonvanishing when these are absent, because the
auxiliary fields transform under local supersymmetry transformations
into gravitino field equations.

The BV formalism involves the construction of an extension $S[\Phi,
\Phi^*]$ of the action $S_0[\Phi]$ that satisfies the master equation
(the superscripts $L$ and $R$ denote left and right differentiation
respectively, where we frequently omit the $L$ for left-derivatives)
 \begin{equation} \label{mastereq}
  (S,S) = - 2 \int\! d^4x\ \frc{\de^R S}{\de\Phi^*_{\!A}}\ \frc{\de^L
  S}{\de\Phi^A} = 0\ .
 \end{equation} 
Here the $\Phi^A$ collectively denote the fields in the supergravity
multiplet and the ghosts, while the $\Phi^*_{\!A}$ are the antifields
conjugate to $\Phi^A$ with respect to the antibracket, i.e.\ $\big(
\Phi^A(x),\Phi^*_B(y)\big)=\de^A_B\,\de(x-y)$. One assigns antifield
numbers ($af$) and ghost numbers ($gh$) to the antifields according to
the relation $af\,\Phi^*_{\!A}=-gh\,\Phi^*_{\!A}=1+gh\,\Phi^A$; for the
fields $af\,\Phi^A=0$. (For example, $af\,A^{*\mu}=1$ and $af\,C_\mu^*=
2$.) Given a solution to the master equation, the corresponding BRST
operator
 \begin{equation}
  s = (S,\cdot\,) = \int\! d^4x\ \Big( \frc{\de^R S}{\de\Phi^A}\
  \frc{\de}{\de\Phi^*_{\!A}} - \frc{\de^R S}{\de\Phi^*_{\!A}}\
  \frc{\de}{\de\Phi^A} \Big)
 \end{equation}
is automatically nilpotent. It decomposes into pieces $s_k$ of definite
antifield number $k$, $s=\sum_{k\geq -1}s_k$, with $s_{-1}\Phi^A=0$
and $s_0\Phi^A$ as in \eqref{s_0-sugra} and \eqref{s_0-ghosts}.

The minimal solution to \eqref{mastereq} contains terms bilinear in the
antifields and is given by
 \begin{align}
  S_\mathrm{min} = S_0 - \int\! d^4x\, \big( & s_0 \e{\mu}{a}\,
	e_{a}^{*\,\mu} + s_0 A_\mu\, A^{*\mu} + \psi^{*\mu}\, s_0
	\psi_\mu + s_0 \bpsi_\mu\, \bpsi{}^{*\mu} \notag \\*[-2pt]
  & + s_0 C\, C^* + s_0 C^\mu\, C^*_\mu + s_0 C^{ab}\, C^*_{ab} -
	\xi^* s_0 \xi + s_0 \bxi\, \bxi{}^* \notag \\*[3pt]
  & + \psi^{*\mu} M_{\mu\nu} \bpsi{}^{*\nu} + \half \psi^{*\mu}
	N_{\mu\nu} \psi^{*\nu\,t} + \half \bpsi{}^{*\mu\,t}
	N^\dag_{\mu\nu} \bpsi{}^{*\nu} \notag \\*[3pt]
  & + \psi^{*\mu} \Sigma^{ab}_\mu C^*_{ab} + \Sigma^{ab}_\mu{}^\dag
	\bpsi{}^{*\mu} C^*_{ab} \big)\ . \label{Smin}
 \end{align}
The terms in the first line have antifield number 1 (except $S_0$, which
has vanishing antifield number), those in the second and third line have
$af=2$ and the last line has $af=3$. This extended Euclidean action is
hermitean if we define $(\Phi_{\!A}^*)^\dag=-(\Phi^{A\dag})^*$ as usual.

From $S_\mathrm{min}$ we obtain the nilpotent BRST transformations of
$\psi_\mu$ and $C^{ab}$
 \begin{align}
  s \psi_\mu & = (S_\mathrm{min}, \psi_\mu) = - \frc{\de
	S_\mathrm{min}}{\de\psi^{*\mu}} = s_0 \psi_\mu + M_{\mu\nu}
	\bpsi{}^{*\nu} + \psi^{*\nu} N_{\nu\mu} + \Sigma^{ab}_\mu
	C^*_{ab} \notag \\[2pt]
  s C^{ab} & = (S_\mathrm{min}, C^{ab}) = - \frc{\de S_\mathrm{min}}{
	\de C^*_{ab}} = s_0 C^{ab} + \psi^{*\mu} \Sigma^{ab}_\mu +
	\Sigma^{ab}_\mu{}^\dag \bpsi{}^{*\mu}\ .
 \end{align}
On all other fields we have $s\Phi^A=s_0\Phi^A$ exactly.
\smallskip

In order to gauge-fix the various local symmetries of the action, we
introduce as usual a nonminimal sector, consisting of Nakanishi-Lautrup
auxiliary fields $b_\mu$, $b_{ab}$, $b$, $\be$, $\be^\dag$, the
antighosts $\bC_\mu$, $\bC_{ab}$, $\bC$, $\bar{\xi}$, $\bar{\xi}^\dag$
(short bars denote antighosts), and the corresponding antifields. To
fix supersymmetry, we employ in addition a second pair of Dirac spinors
$\eta$, $\eta^\dag$ and $\varrho$, $\varrho^\dag$, whose role will be
explained below. $\bC$, $\bC_\mu$, $\bC_{ab}$, $\be$, $\be^\dag$, $\eta$
and $\eta^\dag$ are anticommuting, the other fields are commuting.

The action for the nonminimal sector is
 \begin{equation}
  S_\mathrm{non} = \int\! d^4x\, \big( - b_\mu\, \bC^{*\mu} - b_{ab}\,
  \bC^{*ab} - b\, \bC^* + \bar{\xi}^* \be - \be^\dag \bar{\xi}^{\dag *}
  + \varrho^* \eta - \eta^\dag \varrho^{\dag *} \big)\ .
 \end{equation}
Because the fields and antifields in $S_\mathrm{non}$ are different from
the fields and antifields in $S_\mathrm{min}$, $S=S_\mathrm{min}+
S_\mathrm{non}$ satisfies the master equation \eqref{mastereq}. The BRST
transformations for the nonminimal sector are obtained by taking the
antibracket with $S$,
 \begin{gather}
  s \bC_\mu = b_\mu\ ,\quad s b_\mu = 0 \notag \\[2pt]
  s \bC_{ab} = b_{ab}\ ,\quad s b_{ab} = 0 \notag \\[2pt]
  s \bC = b\ ,\quad s b = 0 \notag \\[2pt]
  s \bar{\xi} = \be\ ,\quad s \be = 0\ ,\quad s \varrho = \eta\ ,\quad
	s \eta = 0 \notag \\*[2pt]
  s \bar{\xi}^\dag = \be^\dag\ ,\quad s \be^\dag = 0\ ,\quad s
	\varrho^\dag = \eta^\dag\ ,\quad s \eta^\dag = 0\ .
	\label{s-antigh}
 \end{gather}
The reason for the last two definitions will become clear below.

We now replace the antifields by
 \begin{equation} \label{af-psi}
  \Phi^*_{\!\cA} = - \frc{\de}{\de\Phi^\cA}\, \Psi[\Phi]\ ,
 \end{equation}
where $\cA$ refers to all fields including the nonminimal sector and
$\Psi$ is a suitably chosen gauge-fixing fermion of ghost number $-1$.
Upon elimination of the $\Phi^*_{\!\cA}$ in $S=S_\mathrm{min}+
S_\mathrm{non}$ we obtain
 \begin{equation}
  S_\Psi[\Phi] = S[\Phi,\Phi^*=-\de\Psi/\de\Phi] = S_0[\Phi] +
  s_0 \Psi + \ldots .
 \end{equation}
The ellipses denote four-ghost terms that arise from the antifield
bilinears in $S$ and will be given explicitly below. $S_\Psi$ is
invariant under the gauge-fixed BRST transformations
 \begin{equation}
  s_\Psi\, \Phi^\cA = s \Phi^\cA|_{\Phi^*=-\de\Psi/\de\Phi}\ .
 \end{equation}
At this point we have specified the quantum action for any choice of
$\Psi$.

We now turn to a suitable choice for $\Psi$.
If one considers a complex gravitino in a background gravitational
field, a gauge-fixing term for local supersymmetry that manifestly
preserves the Euclidean space symmetries is given by $e\,\bpsi\!\cdot\!
\ga\,\ga_5\Ds\ \ga\!\cdot\!\psi$. This gauge-fixing term is obtained
from the usual one in Minkowski spacetime \cite{vN2} by the Wick
rotation discussed above. It is well-known that such a gauge-fixing
term which depends on the vielbein field leads to additional
Nielsen-Kallosh (NK) ghosts \cite{NK}. Since we are considering a
dynamical gravitational field and want to avoid spurious vertices, we
prefer a vielbein-independent gauge-fixing term $\bpsi\!\cdot\!\gah\,
\ga_5\ds\,\gah\!\cdot\!\psi$ with $\gah^\mu\equiv\ga^a\de_a^\mu$ and
$\ds\equiv\gah^\mu\p_\mu$. Both terms give the same propagators.
However, we find it worthwile first to consider the more general case
$e\,\bpsi\!\cdot\!\ga\,\ga_5\Ds\ \ga\!\cdot\!\psi$ and afterwards to
take the flat space limit.

In a path integral approach we could start from
 \begin{equation}
  \Delta_\mathrm{FP}\, \de(\ga\!\cdot\!\psi - \chi)\, \de(\bpsi\!\cdot
  \!\ga - \chi^\dag) \int [d\chi]\, [d\chi^\dag]\, \exp \Big( -
  \frc{1}{2k} \int\! d^4x\ e\, \chi^\dag \ga_5 \Ds\, \chi \Big)\,
  (\det \ga_5 \Ds\,)^{-1}\ .
 \end{equation}
The term $(\det\ga_5\Ds\,)^{-1}$ is a normalization factor of the path
integral over $\chi$ and $\chi^\dag$ and leads to a complex commuting
NK ghost. (For $N=1$ theories one has Majorana spinors $\chi$ and a
factor $(\det\ga_5\Ds\,)^{-1/2}$. One replaces this by $(\det\ga_5
\Ds\,)^{1/2}\,(\det\ga_5\Ds\,)^{-1}$ and obtains then one real
anticommuting NK ghost and one pair of commuting NK ghosts. In Euclidean
space one can take complex holomorphic NK ghosts, similar to the
approach for Majorana spinors in Euclidean space discussed in the
introduction. For $N=2$ theories we can directly take one complex
commuting ghost because the Dirac action for a complex commuting spinor
is nonvanishing.) The factor $\Delta_\mathrm{FP}$ yields the susy ghost
action of the form $\k^{-1}e\bar{\xi}^\dag\ga_5\Ds\ \xi +\text{h.c.}$
(obtained from the usual ghost action in Minkowski spacetime by the Wick
rotation).

We must now find the corresponding expressions in the BV formalism. This
is surprisingly complicated. The most straightforward way to proceed
would be to start from the following nonlocal gauge-fixing fermion
$\Psi$,
 \begin{gather}
  \Psi = \int\! d^4x\ e\, \bar{\xi}^\dag \ga_5 (\ga\! \cdot\! \psi - k\,
	\Ds\,{}^{-1} \be) - \text{h.c.} \notag \\
  s_0 \Psi = \int\! d^4x\ e \big[ \be^\dag \ga_5 (\ga\! \cdot\! \psi -
	k\, \Ds\,{}^{-1} \be) + \bar{\xi}^\dag \ga_5\, s_0 (\ga\! \cdot
	\! \psi) \big] + \text{h.c.}
 \end{gather}
where $k$ is the gauge-fixing parameter. (We are considering a
background gravitational field and a gravitino-independent spin
connection for the time being, so the vielbein fields in the
gauge-fixing fermion are not varied.)
Eliminating $\be$ we would indeed find the desired gauge-fixing term
$e\,\bpsi\!\cdot\!\ga\,\ga_5\Ds\ \ga\!\cdot\!\psi$, \emph{and} the
complex NK ghost from the path integral over $\be$ and $\be^\dag$.
However, the main virtue of the BV formalism is that it tries to avoid
problems connected to the path integral measure, and therefore it is
preferable to extend the above approach in such a way that all nonlocal
terms in the action cancel. This is possible for $N=1$ theories if one
introduces two contractible pairs $\eta_1$, $\varrho_1$ and $\eta_2$,
$\varrho_2$ (all of which are Majorana spinors) with BRST
transformations
 \begin{equation}
  s \eta_1 = \varrho_1\ ,\quad s \varrho_1 = 0\ ,\quad s \varrho_2 =
  \eta_2\ ,\quad s \eta_2 = 0\ .
 \end{equation}
The $\eta_i$ are anticommuting, while the $\varrho_i$ are commuting.
One may add to the action the BRST exact term
 \begin{align}
  s (\eta_1^t \cC \Ds\, \varrho_2) & = \varrho_1^t \cC \Ds\, \varrho_2
	- \eta_1^t \cC \Ds\, \eta_2 \notag \\
  & = \varrho_1^t \cC \Ds\, \varrho_2 - \half \la^t \cC \Ds\, \la +
	\half \chi^t \cC \Ds\, \chi - \half D_\mu (\eta_1^t \cC \ga^\mu
	\eta_2)\ ,
 \end{align}
where $\cC$ is the Euclidean charge conjugation matrix\footnote{The
Euclidean $\cC$ is obtained from the Minkowski matrix $\cC_M$ via $\cC=
O^t\cC_MO$ with $O$ as in \eqref{O} and satisfies the same relations in
Euclidean space as in Minkowski spacetime, namely $\cC^t=-\cC$ and $\cC
\ga_\mu=-\ga_\mu^t\cC$, c.f.\ \cite{vNW}.} and $\la=(\eta_1+\eta_2)/\2$,
$\chi=(\eta_1-\eta_2)/\2$. Shifting $\chi\rightarrow\chi+\Ds\,{}^{-1}
\be$, we find the anticommuting NK ghost $\la$, the two commuting NK
ghosts $\rho_i$, and further a term $\be^t\cC\Ds\,{}^{-1}\be$ which
cancels the nonlocal contribution from the BRST variation of the
gauge-fixing fermion. One is left with terms of the form $\be^t\cC(\ga
\!\cdot\!\psi+2k\chi)+\chi^t\cC\Ds\,\chi$, and integration over $\be$
and $\chi$ yields the desired gauge-fixing term $(\ga\cdot\psi)^t
\cC\Ds\,\ga\cdot\psi$. In this way we end up with the same result
as from the path integral.

We now generalize these results to the case $N=2$, but at the same time
we aim for $\bpsi\!\cdot\!\gah\,\ga_5\ds\,\gah\cdot\!\psi$ as the
supersymmetry fixing term. We have been able to achieve this with only
one new complex contractible pair of fields $\eta$ and $\varrho$.
Consider the following gauge-fixing fermion
 \begin{align}
  \Psi = \int\! d^4x\, \big[ & \bC \big( \p_\mu A_\mu - k_1\, b
	\big) + \bC^\mu \big( \p_\nu (e g^{\mu\nu}) - k_2\, b^\mu
	\big) \notag \\*[-2pt]
  & + \bC^{ab} \big( (\de_a^\mu e_{\mu b} - \de_b^\mu e_{\mu a}) -
	k_3\, b_{ab} \big) + k_4\, \eta^\dag \ga_5 \ds \varrho - k_4\,
	\varrho^\dag \ga_5 \ds \eta \notag \\*
  & - \bar{\xi}^\dag \ga_5 \big( \gah \!\cdot\! \psi - k_4\, \ds^{-1}
	\be \big) + \big( \bpsi \!\cdot\! \gah - k_4\, \be^\dag
	\overset{\leftarrow}{\ds}{}^{-1} \big) \ga_5 \bar{\xi}\,
	\big]\ .
 \end{align}
We chose to contract world indices with $\de_{\mu\nu}$ rather than the
metric in order to have a minimal number of vertices to deal with when
we check the Ward identities. Upon elimination of the $\Phi^*_{\!\cA}$
by using \eqref{af-psi} in $S=S_\mathrm{min}+S_\mathrm{non}$, we obtain
 \begin{align}
  S_\Psi = S_0 + s_0 \Psi + \int\! d^4x\, \big( & \bar{\xi}^\dag \ga_5
	\gah^\mu M_{\mu\nu} \gah^\nu \ga_5 \bar{\xi} - \half\,
	\bar{\xi}^\dag \ga_5 \gah^\mu N_{\mu\nu} (\bar{\xi}^\dag \ga_5
	\gah^\nu)^t \notag \\*[-2pt]
  & - \half\, (\gah^\mu \ga_5 \bar{\xi})^t N^\dag_{\mu\nu} \gah^\nu
	\ga_5 \bar{\xi}\, \big)\ . \label{S}
 \end{align}
The first two lines of \eqref{Smin} have given a result which can also
be written as $s_0\Psi$ by using the chain rule, the third line of
\eqref{Smin} gives the remaining terms in \eqref{S}, while the last
line in \eqref{Smin} vanishes because $\Psi$ does not depend on the
SO(4) ghosts $C^{ab}$. The BRST variation of $\Psi$ yields the
gauge-fixing terms and the ghost action (up to the four-ghost terms in
\eqref{S} which are due to the non-closure of the supersymmetry
algebra),
 \begin{align}
  s_0 \Psi = \int\! d^4x\, \Big[ & \, b\, (\p_\mu A_\mu - k_1\, b) +
	b^\mu \big( \p_\mu (e g^{\mu\nu}) - k_2\, b^\mu \big) +
	b^{ab}\, \big( (e_{ab} - e_{ba}) - k_3\, b_{ab} \big) \notag
	\\*[-2pt]
  & + \p_\mu \bC\, s_0 A_\mu + \p_\mu \bC^\nu s_0 (e g^{\mu\nu}) -
	\bC^{ab} s_0 (e_{ab} - e_{ba}) \notag \\*[2pt]
  & - \be^\dag \ga_5 \gah \!\cdot\! \psi - \bpsi \!\cdot\! \gah \ga_5
	\be + 2 k_4\, \be^\dag \ga_5 \ds^{-1} \be - 2 k_4\, \eta^\dag
	\ga_5 \ds \eta \notag \\*
  & - \bar{\xi}^\dag \ga_5\, s_0 (\gah \!\cdot\! \psi) + s_0 (\bpsi
	\!\cdot\! \gah) \ga_5 \bar{\xi}\, \Big]\ ,
 \end{align}
where we have integrated by parts. The nonlocal term $\be^\dag\ga_5
\ds^{-1}\be$ can be removed by means of a shift $\eta\rightarrow\eta
-\ds^{-1}\be$, after which the third line in the above equation
reads
 \begin{equation}
  - \be^\dag \ga_5 \big( \gah \!\cdot\! \psi - 2 k_4\, \eta \big)
  - \big( \bpsi \!\cdot\! \gah - 2 k_4\, \eta^\dag \big) \ga_5 \be -
  2 k_4\, \eta^\dag \ga_5 \ds \eta\ .
 \end{equation}
We can now eliminate the auxiliary fields by their algebraic equations
of motion, which results in
 \begin{align}
  s_0 \Psi = \int\! d^4x\, \Big[ & \, \frc{1}{4k_1}\, (\p_\mu A_\mu)^2
	+ \frc{1}{4k_2}\, \big( \p_\mu (e g^{\mu\nu}) \big)^2 +
	\frc{1}{4k_3}\, (e_{ab} - e_{ba})^2 \notag \\*[2pt]
  &  + \p_\mu \bC\, s_0 A_\mu + \p_\mu \bC^\nu s_0 (e g^{\mu\nu}) -
	\bC^{ab} s_0 (e_{ab} - e_{ba}) \notag \\*[2pt]
  & - \frc{1}{2k_4}\, \bpsi \!\cdot\! \gah\, \ga_5 \ds (\gah \!\cdot\!
	\psi) - \bar{\xi}^\dag \ga_5\, s_0 (\gah \!\cdot\! \psi) + s_0
	(\bpsi \!\cdot\! \gah) \ga_5 \bar{\xi}\, \Big]\ , \label{sPsi}
 \end{align}
and the BRST transformations of the antighosts turn into
 \begin{gather}
  s_\Psi\, \bC = \frc{1}{2k_1}\, \p_\mu A_\mu\ ,\quad s_\Psi\, \bC^\mu
	= \frc{1}{2k_2}\, \p_\nu (e g^{\mu\nu}) \notag \\
  s_\Psi\, \bC^{ab} = \frc{1}{2k_3}\, (e^{ab} - e^{ba})\ ,\quad s_\Psi\,
	\bar{\xi} = \frc{1}{2k_4}\, \ds (\gah \!\cdot\! \psi)\ .
 \end{gather}
At this point the complete quantum action is explicitly known: it is
given by \eqref{S} and \eqref{sPsi}.

Our aim is to explicitly check the infinite and finite parts of Ward
identities. To avoid the complications of propagators in anti-de~Sitter
space, we set the cosmological constant $g=0$ (the present paper is a
pilot program for more complicated studies with $g\neq0$). Hence we
expand about flat space. A convenient choice for the gauge-fixing
parameters is $k_1=\half$, $k_2=\k^2$, $k_3=0$, $k_4=1$, which yields
the following propagators for the supergravity multiplet
 \begin{gather}
  \grn{A_\mu\, A_\nu}_0 = \frc{1}{p^2}\, \de_{\mu\nu}\ ,\quad \grn{
	c_{\mu\nu}\, c_{\rho\si}}_0 = \frc{1}{2p^2}\, (\de_{\mu\rho}
	\de_{\nu\si} + \de_{\mu\si} \de_{\nu\rho} - \de_{\mu\nu}
	\de_{\rho\si}) \notag \\[2pt]
  \grn{\psi_\mu\, \bpsi_\nu}_0 = \frc{\i}{2p^2}\, \ga_5 \gah_\nu \ps\,
	\gah_\mu\ ,
 \end{gather}
where $\e{\mu}{a}=\de_\mu^a+\k\,c_\mu{}^a$. The ghost propagators read
 \begin{gather}
  \grn{C\,\bC}_0 = \frc{1}{p^2}\ ,\quad \grn{C^\mu\,\bC^\nu}_0 = -
	\frc{1}{p^2}\, \de^{\mu\nu} \notag \\[2pt]
  \grn{\hat{C}_{ab}\,\bC^{cd}}_0 = \frc{1}{2}\, \de_{[a}^c \de_{b]}^d\
	,\quad \grn{\xi\,\bar{\xi}^\dag}_0 = \k \frc{\i}{p^2}\, \ga_5\,
	\ps\ ,
 \end{gather}
where we have shifted $\hat{C}_{ab}=C_{ab}+\de_{\mu[a}\p_{b]}C^\mu$ in
order to diagonalize the kinetic terms of the ghosts \cite{vN2}. Note
the occurence of $\ga_5$ in the propagators of the spinors.
\medskip

We are now ready to compute Ward identities. These are derived by first
making the substitution $\Phi^*_{\!\cA}=K_\cA-\de\Psi/\de\Phi^\cA$
instead of \eqref{af-psi}, where the $K_\cA$ are a set of external
sources for the gauge-fixed BRST transformations\footnote{One has
$S_\Psi[\Phi,K]=S_0+s_0\Psi-\int s_\Psi\Phi^\cA\,K_\cA+\dots$.}
introduced in \cite{ZJ}. Since this substitution is a canonical
transformation, $S_\Psi[\Phi,K]$ satisfies a master equation similar
to \eqref{mastereq}
 \begin{equation}
  \int\! d^4x\ \frc{\de^R S_\Psi}{\de K_\cA}\ \frc{\de^L S_\Psi}{\de
  \Phi^\cA} = 0\ .
 \end{equation}
Now consider the generating functional
 \begin{equation}
  Z[J,K] = \int [d\Phi]\, \exp \Big( - S_\Psi[\Phi,K] + \int\! d^4x\
  \Phi^\cA J_\cA \Big)\ ,
 \end{equation}
and perform an infinitesimal change of integration variables $\Phi^\cA
\rightarrow\Phi^\cA+\de^R S_\Psi/\de K_\cA$. Assuming the path integral
measure to be invariant, one obtains
 \begin{equation}
  \int\! d^4x\ \frc{\de^R Z}{\de K_\cA}\, J_\cA = 0\ .
 \end{equation}
The same relation holds for the generating functional $W=\ln Z$ of
connected Green functions. (Right-) differentiation with respect to the
sources $J_\cA$ and setting $J_\cA=K_\cA=0$ afterwards then yields the
Ward identities for connected graphs
 \begin{equation} \label{s_grn}
  s_\Psi \grn{\Phi^{\cA_1}(x_1) \dots \Phi^{\cA_n}(x_n)} = 0\ ,
 \end{equation}
expressing the BRST invariance of the Green functions.

We consider the identity for the two-point function $\grn{A_\mu(x)\,
\bC(y)}$. The Fourier transformed identity reads
 \begin{align}
  0 & = - \i p_\nu \grn{A_\mu(p)\, A_\nu(-p)} + \i p_\mu \grn{C(p)\,
	\bC(-p)} \notag \\*
  & \tab + \grn{C^\nu\! * \p_\nu A_\mu(p)\, \bC(-p)} + \grn{\p_\mu
	C^\nu\! *\! A_\nu(p)\, \bC(-p)} \notag \\*
  & \tab + \tfrac{\i}{\2}\, \grn{\bxi \ga_5 * \psi_\mu(p)\, \bC(-p)} -
	\tfrac{\i}{\2}\, \grn{\bpsi_\mu \ga_5 * \xi(p)\, \bC(-p)}\ .
 \end{align}
It obviously holds at tree level. For $g=0$ the relation is easily
verified also at the one-loop level. From \eqref{S} and \eqref{sPsi} we
infer that there are no vertices involving the ghost $C$, so only
the Green function $\grn{A_\mu\,A_\nu}$ receives any loop contributions
at all and the Ward identity expresses the transversality of the sum of
these graphs. The one-loop diagrams consist of a gravitino loop, a
vector-graviton loop and various tadpoles (i.e., loops with one internal
line). The latter vanish in dimensional regularization since all fields
are massless. Transversality of $\grn{A_\mu\,A_\nu}$ at one loop now
follows from the fact that $A_\mu$ enters the relevant vertices only via
its field strength, which implies that the contraction with $p_\nu$
vanishes. Of course, this argument hinges on a suitable regularization
that can deal with the presence of $\ga_5$ and the Levi-Civita tensor in
the vector-gravitino vertices.

This Ward identity is admittedly very simple. Other identities are quite
complicated due to the gravitational interactions. Our aim was to
develop the formalism to the point of explicit Feynman diagrams, and
this we have achieved.

\section{Euclidean N=2 Super Yang-Mills} 

As explained in the introduction, the continuation to Euclidean space of
theories which contain pseudoscalars gives rise to certain unusual signs
and factors of $\i$ in the action and transformation rules that
originate from the Wick rotation $\varphi\rightarrow\i\varphi_E$ of
pseudoscalars $\varphi$. For an ordinary scalar $\phi$ one has of course
just $\phi\rightarrow\phi_E$. In particular, the kinetic terms of the
pseudoscalars enter the Euclidean action with the ``wrong'' sign, which
leads to propagators with a sign opposite to that of the scalar
propagators. At first sight this seems to spoil the Ward identities,
since diagrams containing internal pseudoscalar lines acquire a factor
$(-1)$ from each of the corresponding propagators. However, the vertices
also acquire a factor $\i$ for each pseudoscalar, leading to an
additional factor $(-1)$ from the two endpoints of every internal
pseudoscalar line. These restore the signs of the propagators, such that
there is no violation of the Ward identities coming from the Wick
rotation of the pseudoscalars.

We study the Ward identities in some detail in the example of $N=2$
supersymmetric Yang-Mills theory. The on-shell field content consists
of one vector $A_\mu^I$, one scalar $\phi^I$, one pseudoscalar
$\varphi^I$, and one Dirac spinor $\la^I$ for each generator $T_I$ of
the gauge group. The classical action in Euclidean space has been given
first by Zumino in \cite{Z} and coincides with the one derived from the
Wick rotation rules of \cite{vNW}. It reads (dropping again all
subscripts $E$)
 \begin{align}
  S_0 = \int\! d^4x\, \Big[\, & \frc{1}{4}\, F^{\mu\nu I} F_{\mu\nu}^I
	+ \frc{1}{2}\, D^\mu \phi^I\, D_\mu \phi^I - \frc{1}{2}\, D^\mu
	\varphi^I\, D_\mu \varphi^I - \bla{}^I \ga_5 \ga^\mu D_\mu
	\la^I \notag \\*
  & - \i g f^{IJK} \bla{}^I (\ga_5 \phi^J + \varphi^J) \la^K -
	\frc{1}{2}\, (g f^{IJK} \phi^J \varphi^K)^2 \Big]\ ,
 \end{align}
with covariant derivative $D_\mu\phi^I=\p_\mu\phi^I+gf^{IJK}\!A_\mu^J
\phi^K$, etc. Note the signs of the kinetic term of $\varphi^I$ and
of the scalar potential; clearly, the Euclidean action is not bounded
from below. For this model a set of three real scalar auxiliary fields
is known, but we shall not use them for reasons explained in the
introduction.

The BRST transformations in Minkowski space in absence of auxiliary
fields in the supersymmetry multiplet have been worked out in \cite{M}.
Using the Wick rotation rules of \cite{vNW} one finds for the
antifield-independent parts of the Euclidean BRST transformation rules
 \begin{align}
  s_0 A_\mu^I & = D_\mu C^I + \i\, \bxi \ga_5 \ga_\mu \la^I + \i\,
	\bla{}^I \ga_5 \ga_\mu \xi + C^\nu \p_\nu A_\mu^I \notag
	\\[2pt]
  s_0 \phi^I & = - g f^{IJK} C^J \phi^K - \bxi \ga_5 \la^I - \bla{}^I
	\ga_5 \xi + C^\mu \p_\mu \phi^I \notag \\[2pt]
  s_0 \varphi^I & = - g f^{IJK} C^J \varphi^K + \bxi \la^I + \bla{}^I
	\xi + C^\mu \p_\mu \varphi^I \notag \\[2pt]
  s_0 \la^I & = - g f^{IJK} C^J \la^K - D_\mu (\phi^I - \varphi^I \ga_5)
	\ga^\mu \xi + \tfrac{\i}{2}\, F_{\mu\nu}^I\, \ga^{\mu\nu} \xi
	\notag \\*
  & \tab + \i g f^{IJK} \phi^J \varphi^K\, \ga_5 \xi + C^\mu \p_\mu
	\la^I \notag \\[2pt]
  s_0 \bla{}^I & = - g f^{IJK} C^J \bla{}^K + \bxi \ga^\mu D_\mu (\phi^I
	- \ga_5 \varphi^I) - \tfrac{\i}{2}\, \bxi \ga^{\mu\nu} F_{\mu
	\nu}^I \notag \\*
  & \tab + \i g f^{IJK} \phi^J \varphi^K\, \bxi \ga_5 + C^\mu \p_\mu
	\bla{}^I\ . \label{s0_SYM}
 \end{align}
Here the $C^I$ are the ghosts of gauge transformations. Further, $\xi$,
$\bxi$ and $C^\mu$ denote again the supersymmetry and translation ghosts
respectively, only now these are constant since we are considering rigid
supersymmetry and translations.

The BRST transformations of the ghosts read
 \begin{align}
  s_0 C^I & = - \half g f^{IJK} C^J C^K + 2\i\, \bxi (\ga_5 \phi^I +
	\varphi^I) \xi + 2\, \bxi \ga_5 \ga^\mu \xi\, A_\mu^I + C^\mu
	\p_\mu C^I \notag \\[2pt]
  s_0 C^\mu & = - 2\, \bxi \ga_5 \ga^\mu \xi \notag \\[2pt]
  s_0 \xi & = 0\ . \label{s0_gh}
 \end{align}

As in the supergravity case, the BRST operator $s_0$ is not nilpotent.
We have
 \begin{equation}
  s_0^2\, \la^I = M\, \frc{\de S_0}{\de \bla{}^I} + \frc{\de S_0}{\de
  \la^I}\, N\ ,\quad s_0^2\, \text{(other fields)} = 0\ ,
 \end{equation}
where
 \begin{equation}
  M^\al{}_\be = 2 (\bxi \xi) \de^\al_\be - (\ga_5 \xi)^\al\, (\bxi
  \ga_5)_\be - (\ga^\mu \ga_5 \xi)^\al\, (\bxi \ga_5 \ga_\mu)_\be\
  ,\quad N^{\al\be} = - \xi^\al\, \xi^\be\ .
 \end{equation}
A nilpotent BRST operator $s$ can be obtained from a solution $S$ of
an extended master equation \cite{BC,BHW}
 \begin{equation}
  (S,S) = - 2 \int\! d^4x\, \Big( \frc{\de^R S}{\de\Phi^*_{\!A}}\
  \frc{\de^L S}{\de\Phi^A} \Big) - 2\, \frc{\p^R S}{\p\mathcal{C}^*_r}\
  \frc{\p^L S}{\p\mathcal{C}^r} = 0\ .
 \end{equation}
Here the fields $\Phi^A$ consist of the supersymmetry multiplet and the
ghosts $C^I$ (and later also the BRST auxiliary fields and antighosts),
while the $\mathcal{C}^r$ denote the constant ghosts $C^\mu$, $\xi$,
$\bxi$ of the rigid symmetries. The minimal solution is given by
 \begin{align}
  S_\mathrm{min} = S_0 - \int\! & d^4x\, \big( s_0 A_\mu^I\, A^{*\mu I}
	+ s_0 \phi^I\, \phi^{*I} + s_0 \varphi^I\, \varphi^{*I}
	+ \la^{*I}\, s_0 \la^I + s_0 \bla{}^I\, \bla{}^{*I} \notag
	\\*[-2pt]
  & + s_0 C^I\, C^{*I} + \la^{*I} M \bla{}^{*I} - \half (\la^{*I}
	\xi)^2 - \half (\bxi \bla{}^{*I})^2 \big) + 2\, \bxi \ga_5
	\ga^\mu \xi\, C^*_\mu\ .
 \end{align}
This action is hermitean because $(\Phi_{\!A}^*)^\dag=-(\Phi^{A
\dag})^*$. The only constant term in this action is the last one. It
follows that $s=(S_\mathrm{min},\cdot\,)=s_0$ on all fields $\Phi^A$
and constant ghosts $\mathcal{C}^r$ except for the gauginos $\la^I$,
for which we find
 \begin{equation}
  s \la^I = s_0 \la^I + 2 (\bxi \xi)\, \bla{}^{*I} - (\bxi \ga_5
  \bla{}^{*I})\, \ga_5 \xi + (\bxi \ga_5 \ga_\mu \bla{}^{*I})\, \ga_5
  \ga^\mu \xi - (\la^{*I} \xi)\, \xi\ .
 \end{equation}

In the gauge-fixing procedure we follow \cite{M} and include also a
translation in the BRST transformations of the antighosts $\bC^I$.
(In \eqref{s0_SYM} and \eqref{s0_gh} there are already translations
of the classical fields and internal symmetry ghosts.)
The master equation then implies that the auxiliary fields are not
BRST-invariant, for the solution for the nonminimal sector now reads
 \begin{equation}
  S_\mathrm{non} = - \int\! d^4x\, \big( s \bC^I\, \bC^{*I} + s b^I\,
  b^{*I} \big)\ ,
 \end{equation}
where
 \begin{equation} \label{sbC}
  s \bC^I = b^I + C^\mu \p_\mu \bC^I\ ,\quad s b^I = 2\, \bxi \ga_5
  \ga^\mu \xi\, \p_\mu \bC^I + C^\mu \p_\mu b^I\ .
 \end{equation}
The term $s\bC^I=b^I$ is standard, but adding the translation term
$C^\mu\p_\mu\bC^I$ BRST nilpotency leads to a nonvanishing
transformation of $b^I$. As gauge-fixing fermion for the local symmetry
we choose
 \begin{equation} \label{PsiSYM}
  \Psi = \int\! d^4x\, \bC^I \big( \p_\mu A^{\mu I} - k\, b^I
  \big)\ ,
 \end{equation}
where the real constant $k$ denotes again the gauge-fixing parameter.
Upon replacing $\Phi^*_{\!\cA}=-\de\Psi/\de\Phi^\cA$ in $S=S_\mathrm{%
min}+S_\mathrm{non}$ the terms with two antifields vanish, whereas the
terms with one antifield yield the BRST variation of $\Psi$
 \begin{align}
  s_0 \Psi = s \Psi = \int\! d^4x\, \big[ & b^I (\p_\mu A^{\mu I} - k\,
	b^I) + \p_\mu \bC^I \big( D^\mu C^I - 2k\, \bxi \ga_5 \ga^\mu
	\xi\, \bC^I \notag \\*[-2pt]
  & + \i\, \bxi \ga_5 \ga^\mu \la^I + \i\, \bla{}^I \ga_5 \ga^\mu \xi
	\big) \big]\ . \label{sPsiSYM}
 \end{align}
The final quantum action is given by $S_\Psi[\Phi,\xi,\bxi]=S_0[\Phi]+s_0
\Psi$. The result for the gauge-fixing and ghost action is independent
of the translation ghosts because we included translations in
\eqref{sbC} (as a result, $\Psi$ in \eqref{PsiSYM} is translation
invariant). After elimination of the $b^I$ the transformations of the
antighosts read
 \begin{equation} \label{s_PsibC}
  s_\Psi \bC^I = \frc{1}{2k}\, \p_\mu A^{\mu I} + C^\mu \p_\mu \bC^I\ .
 \end{equation}
The final BRST transformation rules $s_\Psi\Phi^\cA$ are then given by
\eqref{s0_SYM}, \eqref{s0_gh} and \eqref{s_PsibC}.

The opposite sign of the kinetic terms of the pseudoscalars $\varphi^I$
implies an opposite sign in the propagators. They read for the bosonic
fields (taking $k=\half$)
 \begin{equation} \label{grnSYM1}
  \grn{A_\mu^I\, A_\nu^J}_0 = \frc{1}{p^2}\, \de_{\mu\nu} \de^{IJ}\
  ,\quad \grn{\phi^I\, \phi^J}_0 = \frc{1}{p^2}\, \de^{IJ}\ ,\quad
  \grn{\varphi^I\, \varphi^J}_0 = - \frc{1}{p^2}\, \de^{IJ}\ .
 \end{equation}
The fields $\Phi^\cA$ we couple to external sources as usual, but the
constant ghosts $\mathcal{C}^r$ are treated themselves as external
sources (so the generating functional $Z$ depends on $\mathcal{C}^r$.)
We have then a nondiagonal kinetic matrix due to the mixed terms in
\eqref{sPsiSYM}
 \begin{gather}
  \int\! d^4x\, \big[ - \bla{}^I \ga_5 \ds \la^I + \p_\mu \bC^I (\p^\mu
	C^I - 2k\, \bxi \ga_5 \ga^\mu \xi\, \bC^I + \i\, \bxi \ga_5
	\ga^\mu \la^I + \i\, \bla{}^I \ga_5 \ga^\mu \xi) \big] =
	\notag \\[-2pt]
  = \frc{1}{2} \int\! \frc{d^4p}{(2\pi)^4} \begin{pmatrix} - \bC^I(p)\,
	, &\! C^I(p)\, , &\! \2\, \bla{}^I(p) \end{pmatrix} \mathcal{M
	}^{IJ} \begin{pmatrix} \bC^J(-p) \\[2pt] C^J(-p) \\[2pt] \2\,
	\la^J(-p) \end{pmatrix}\ ,
 \end{gather}
where $\mathcal{M}^{IJ}$ is given by
 \begin{equation} \label{M^IJ}
  \mathcal{M}^{IJ} = \de^{IJ} \begin{pmatrix} \ 4\i k\, \bxi \ga_5 \ps
  \xi & - p^2 & \2\, \bxi \ga_5 \ps\ \\[2pt] - p^2 & 0 & 0 \\[2pt]
  - \2\, \ga_5 \ps \xi & 0 & \i \ga_5 \ps \end{pmatrix}\ .
 \end{equation}
This matrix is easily inverted. The result is particularly simple in the
Feynman gauge $k=\half$, which we had adopted already above,
 \begin{equation}
  (\mathcal{M}^{-1})^{IJ} = \frc{\i}{p^2}\, \de^{IJ} \begin{pmatrix}
  \ 0 & \i & 0 \\[2pt] \ \i & (2 - 4k)\, \bxi \ga_5 \ps \xi / p^2 &
  - \2\, \bxi \\[2pt] \ 0 & \2\, \xi & \ga_5 \ps \end{pmatrix}\ .
 \end{equation}
With $k=\half$ we find the following nonvanishing propagators
 \begin{equation} \label{grnSYM2}
  \grn{\la^I\, \bla{}^J}_0 = - \frc{\i}{p^2}\, \ga_5 \ps\, \de^{IJ}\
  ,\quad \grn{C^I\, \bC^J}_0 = \frc{1}{p^2}\, \de^{IJ}\ ,\quad \grn{
  \la^I\, C^J}_0 = \frc{\i}{p^2}\, \xi\, \de^{IJ}\ .
\end{equation}
Note that it was the inclusion of translations in the BRST
transformations of the anti\-ghosts that led to the $4\i k\,\bxi\ga_5
\ps\xi$ term in \eqref{M^IJ} and that enabled us to make the $\grn{C^I
\,C^J}_0$ propagator vanish by a suitable choice of the gauge-fixing
parameter. Note also that one may diagonalize the propagators in the
$\la^I\,C^J$ sector by means of a field redefinition $\la^I\rightarrow
\la^I-\i\xi\bC^I$. However, this would result in additional vertices
and more complicated BRST transformations, which we prefer to avoid.

The nonvanishing $\grn{\la^I\,C^J}_0$ propagator signals the
intertwining of gauge and supersymmetry. Rigid supersymmetry is broken
as usual by the gauge-fixing terms, but it fuses with the usual BRST
symmetry for the internal local symmetries into an extended BRST
symmetry which connects the local and the rigid symmetries. This fusion
can also be observed in the Ward identities. These are derived similarly
to the previous case of supergravity, only we set $\mathcal{C}^*_r=0$ in
addition to the substitution $\Phi^*_{\!\cA}=K_\cA-\de\Psi/\de\Phi^\cA$
in $S=S_\mathrm{min}+S_\mathrm{non}$. (We could also have introduced
an external source $K_\mu$ for $C^\mu$ to remove the last term in
\eqref{ME}, but we prefer to treat the constant ghosts as external
fields.) The action $S_\Psi[\Phi,K,\mathcal{C}]=S_0+s_0\Psi-\int s_\Psi
\Phi^\cA\,K_\cA+\dots$ thus obtained then satisfies the master equation
 \begin{equation} \label{ME}
  \int\! d^4x\, \Big( \frc{\de^R S_\Psi}{\de K_\cA}\ \frc{\de^L S_\Psi}{
  \de \Phi^\cA} \Big) + 2\, \bxi \ga_5 \ga^\mu \xi\, \frc{\p S_\Psi}{\p
  C^\mu} = 0\ ,
 \end{equation}
which implies the following Ward identity for the generating functional
$Z[J,K,\mathcal{C}]$
 \begin{equation} \label{WISYM}
  \int\! d^4x\, \Big( \frc{\de^R Z}{\de K_\cA}\, J_\cA \Big) + 2\, \bxi
  \ga_5 \ga^\mu \xi\, \frc{\p Z}{\p C^\mu} = 0\ .
 \end{equation}
We are now interested in the Ward identities \eqref{s_grn}, which are
obtained by differentiating the above equation with respect to the
sources $J_\cA$ and setting $J_\cA=K_\cA= 0$. The second term in
\eqref{WISYM} does not contribute to \eqref{s_grn} since $C^\mu$ enters
$S_\Psi$ only via $\int C^\mu\p_\mu\Phi^\cA\,K_\cA$, which when
differentiated with respect to $C^\mu$ vanishes after setting $K_\cA=0$.

The $\grn{\la^I\,C^J}_0$ propagator is crucial already at tree level.
Consider as an example the Ward identity
 \begin{equation}
  s_\Psi \grn{\la^I(x)\, A_\mu^J(y)} = 0\ .
 \end{equation}
At tree level the composite operators in the BRST transformations do not
contribute and the Fourier transformed relation reads, writing only the
nonvanishing two-point functions,
 \begin{equation}
  \grn{A_\rho^I\, A_\mu^J}_0\, \ga^{\rho\nu} \xi\, p_\nu - \i \grn{
  \la^I\, \bla{}^J}_0\, \ga_5 \ga_\mu \xi + \i p_\mu\, \grn{\la^I\,
  C^J}_0 = 0\ .
 \end{equation}
The first two terms come from the supersymmetry transformations of
$\la^I$ and $A_\mu^J$ respectively, while the last one originates from
the gauge transformation of $A_\mu^J$. The contributions from the
translation ghosts in $s_\Psi\la^I$ and $s_\Psi A_\mu^J$ cancel due to
translational invariance. Plugging in the propagators, we find
 \begin{equation}
  \frc{1}{p^2}\, \de^{IJ}\, (\ga_{\mu\nu} - \ga_5 \ga_\nu \ga_5 \ga_\mu
  - \de_{\mu\nu}) \xi\, p^\nu = 0\ ,
 \end{equation}
which is indeed satisfied. This checks the $\grn{\la^I\,C^J}_0$
propagator in \eqref{grnSYM2}.

In order to investigate the effects of the sign of the pseudoscalar
propagator, we shall consider the following Ward identity for
connected Green functions
 \begin{equation}
  \frc{\p}{\p\xi}\, s_\Psi \grn{\la^I(x)\, \varphi^J(y)}\, |_{\xi=\bxi
  =0} = 0\ .
 \end{equation}
The Fourier transformed identity reads
 \begin{align}
  0 & = \grn{\la^I(p)\, \bla{}^J(-p)} - \i \ga_5 \ps\, \grn{\varphi^I
	(p)\, \varphi^J(-p)} - g f^{IKL} \ga_5 \ga^\mu \grn{A_\mu^K\!
	* \varphi^L(p)\, \varphi^J(-p)} \notag \\*
  & \tab + g \frc{\p}{\p\xi}\, \big[ f^{IKL} \grn{C^K\! * \la^L(p)\,
	\varphi^J(-p)} - f^{JKL} \grn{\la^I(p)\, C^K\! * \varphi^L(-p)}
	\big]\, |_{\xi=\bxi=0}\ , \label{WI_SYM}
 \end{align}
where we have written only terms that contribute up to one-loop order.
Using \eqref{grnSYM1} and \eqref{grnSYM2} this relation is satisfied at
tree level thanks to the nonstandard sign of the pseudoscalar
propagator. At the one-loop level, the gaugino self-energy receives
contributions from virtual $A_\mu^I$, $\phi^I$ and $\varphi^I$, while
the pseudoscalar self-energy gets contributions from a $\la^I$ loop
and a virtual $A_\mu^I$. We regularize the divergent integrals by
dimensional reduction \cite{S}, where only coordinates and momenta are
continued to dimensions $d<4$, while the spinors and vectors remain
four-dimensional. Furthermore, we use $\mathrm{tr}(\ga_\mu\ga_\nu)=4
\de_{\mu\nu}$ and $\{\ga_\mu,\ga_5\}=0$ for all $\mu$, i.e., algebraic
manipulations involving $\ga$-matrices are performed in four dimensions
\cite{TvN2}. We then find for the one-loop Green functions
 \begin{align}
  \grn{\varphi^I(p)\, \varphi^J(-p)} & = 0 \notag \\[2pt]
  \grn{\la^I(p)\, \bla{}^J(-p)} & = 2\, \i \ga_5 \ps\, g^2 f^{IKL}
	f^{JKL}\, I(p^2;d) \notag \\[2pt]
  \grn{A_\mu^K\! * \varphi^L(p)\, \varphi^J(-p)} & = \tfrac{3}{2}\, \i
	p_\mu\, g f^{JKL}\, I(p^2;d) \notag \\[2pt]
  \grn{C^K\! * \la^L(p)\, \varphi^J(-p)} & = \half\, \i \ga_5 \ps \xi\,
	g f^{JKL}\, I(p^2;d) \notag \\[2pt]
  \grn{\la^I(p)\, C^K\! * \varphi^L(-p)} & = \i \ga_5 \ps \xi\, g
	f^{IKL}\, I(p^2;d)\ ,
 \end{align}
where
 \begin{equation*}
  I(p^2;d) = \frc{1}{p^2}\, \frc{1}{(4\pi)^2}\, \Gamma(2-d/2)\,
  \int_0^1\! d\al\, \Big( \frc{4\pi}{\al(1-\al)p^2} \Big)^{2-d/2}\ .
 \end{equation*}
When plugged into \eqref{WI_SYM}, the poles and finite terms cancel and
the Ward identity is verified.

The examples we have investigated show that at the perturbative level
the Ward identities are satisfied if one uses the straightforward
propagators (with extra minus signs for the pseudoscalars). One can
therefore calculate Feynman graphs in Euclidean supersymmetric theories
as easily as in Minkowskian theories.
\bigskip

\parindent 0em

\textbf{\large Acknowledgements} 
\medskip

It is a pleasure to thank Warren Siegel and Mikhail S.~Volkov for useful
discussions.

\end{document}